\documentclass[10pt,a4paper]{article}
\usepackage[left=14mm, top=10mm, right=14mm, bottom=20mm, nohead, foot=10mm]{geometry}
\usepackage[affil-it]{authblk}

\let\OLDthebibliography\thebibliography
\renewcommand\thebibliography[1]{
  \OLDthebibliography{#1}
  \small
  \setlength{\parskip}{0pt}
  \setlength{\itemsep}{0pt plus 0.3ex}
}

\usepackage{multicol}

\usepackage{caption}

\usepackage{misccorr}

\usepackage[english]{babel}
\usepackage{amsmath,amssymb}
\usepackage{textcomp}

\usepackage{xcolor}

\usepackage{hyperref}
\hypersetup{colorlinks=true, linkcolor=[rgb]{0.6 0 0}, citecolor=[rgb]{0 0.4 0}, urlcolor=[rgb]{0 0 0.6}}

\usepackage{graphicx}
\usepackage{epsfig}

\newenvironment{Figure}
  {\par\medskip\noindent\minipage{\linewidth}\captionsetup{type=figure}}
  {\endminipage\par\medskip}
\usepackage{epstopdf}

\title{\large \textbf{Interaction of Individual Skyrmions in Nanostructured Cubic Chiral Magnet}}
\author[1,2]{\normalsize H.~Du}
\author[3]{\normalsize X.~Zhao}
\author[4]{\normalsize F.\,N.~Rybakov}
\author[5]{\normalsize A.\,B.~Borisov}
\author[1]{\normalsize S.~Wang}
\author[1]{\normalsize J.~Tang}
\author[1]{\normalsize C.~Jin}
\author[3]{\normalsize C.~Wang}
\author[1]{\normalsize W.~Wei}
\author[6]{\normalsize N.\,S.~Kiselev\thanks{n.kiselev@fz-juelich.de}}
\author[1,7,8]{\normalsize Y.~Zhang}
\author[3]{\normalsize R.~Che\thanks{rcche@fudan.edu.cn}}
\author[6]{\normalsize S.~Bl\"{u}gel}
\author[1,2,8]{\normalsize M.~Tian\thanks{tianml@hmfl.ac.cn}}

\affil[1]{\small
The Anhui Province Key Laboratory of Condensed Matter Physics at Extreme Conditions, High Magnetic Field Laboratory, Chinese Academy of Sciences and University of Science and Technology of China, Hefei 230026, China}

\affil[2]{\small
Department of Physics, School of Physics and Materials Science, Anhui University, Hefei 230601, China}

\affil[3]{\small
Laboratory of Advanced Materials, Department of Materials Science, Collaborative Innovation Center of Chemistry for Energy Materials, Fudan University, Shanghai 200438, China}

\affil[4]{\small
Department of Physics, KTH-Royal Institute of Technology, Stockholm, SE-10691 Sweden}

\affil[5]{\small
M.\,N.~Miheev Institute of Metal Physics of Ural Branch of Russian Academy of Sciences, Ekaterinburg 620990, Russia}

\affil[6]{\small
Peter Gr\"{u}nberg Institute and Institute for Advanced Simulation, Forschungszentrum J\"{u}lich and JARA, 52425 J\"{u}lich, Germany}

\affil[7]{\small
Department of Physics, University of Science and Technology of China, 230031, China}

\affil[8]{\small
Collaborative Innovation Center of Advanced Microstructures, Nanjing University, Nanjing 210093, China}

\date{}

\begin{document}

%{\let\newpage\relax\maketitle}

\maketitle
\vspace{-3.0\baselineskip}

\renewcommand{\abstractname}{\vspace{-\baselineskip}}
\begin{abstract}
We report the direct evidence of field-dependent character of the interaction between  individual magnetic skyrmions as well as between skyrmions and edges in B20-type FeGe nanostripes observed by means of high resolution Lorentz transmission electron microscopy. 
It is shown that above certain critical values of external magnetic field the character of such long-range skyrmion interactions change from attraction to repulsion. 
Experimentally measured equilibrium inter-skyrmion and skrymion-edge distances as function of applied magnetic field shows quantitative agreement with the results of micromagnetic simulations. 
Important role of demagnetizing fields and internal symmetry of three-dimensional magnetic skyrmions are discussed in details. 
\end{abstract}

%\section*{Introduction}

\begin{multicols}{2}[]

Magnetic chiral skyrmions are topological solitons appearing in  magnetic crystals with broken inversion symmetry~\cite{Bogdanov_89}.
The stability or metastability of such particlelike objects is provided by competition between Heisenberg exchange, Dzyaloshinskii-Moriya interaction (DMI) and interaction with applied field~\cite{Ivanov, Dzyaloshinskii, Moriya, Bogdanov_1995, Yu_10, Nagaosa_Tokura}.
Their small size, topological protection and high mobility make them hold great promise as data bit carriers in novel type of magnetic memory and logical elements for spintronics~\cite{Kiselev11, Fert_13, Zhang_15(1)}.
For the applications as well as for fundamental research, one of the key question concerns the character of skyrmion-skyrmion and skyrmion-edge interactions, which define the equilibrium inter-particle distances and place certain restrictions on the data capacity and ultimate operation speed~\cite{Iwasaki_13,Sampaio_13,Zhang_15(2)}.

Magnetic skyrmions are also known to be stable in ultra thin films and multilayers with strong perpendicular anisotropy and surface/interface induced DMI~\cite{Romming_15, Boulle_16}. 
Such two-dimensional (2D) skyrmions show a long-range interparticle repulsion~\cite{Bogdanov_1995, Bogdanov_11, Zhang_15(2)} and share
a lot of similarities with magnetic bubble domains in perpendicular anisotropy films/multilayers~\cite{Bobeck_75, Hubert_98} and the vortices in type-II superconductors~\cite{Stan}. 
The focus of this work is on the experimental and theoretical study of the inter-skyrmion and skyrmion-edge interactions in cubic chiral magnets with bulk-type DMI. 

Contrary to 2D systems, in cubic chiral magnets the skyrmions are inhomogeneous along the film thickness and form the three-dimensional (3D) skyrmion tubes (SkTs) with extra twist at the ends~\cite{Rybakov_13} due to the effect of chiral surface twist~\cite{Meynell_2014}. 
Moreover, in a wide range of external magnetic field $B_\mathrm{ext}$ the ground state corresponds to a cone phase characterized by chiral modulations of magnetization along $\mathbf{B}_\mathrm{ext}$ and represents non-homogeneous ``vacuum'' for 3D SkTs.
Such SkTs first have been investigated theoretically in~\cite{Rybakov_15} where it was shown that in certain range of parameters the single SkT embedded in the metastable cone phase represents energetically unfavorable state while the global energy minimum corresponds to the lattice of such SkTs. 
This fact indicates that clumping of 3D SkTs may lead to a significant decrease of the total energy. 
Later, by means of numerical methods it was established that asymptotic behavior of 3D SkT possess positive energy density with respect to surrounding cone phase~\cite{Leonov_16}. 
Because of that the inter-particle as well as particle-edge interactions of 3D SkTs should have the character of long-range attraction~\cite{Leonov_APL}
which naturally leads to the formation of skyrmion pairs (dimers) and large skyrmion clusters.
The observation of such clusters have been reported earlier in nanostructured B20-type FeGe~\cite{Du_15, Zhao_16} and extended plates of Cu$_{2}$OSeO$_{3}$~\cite{Muller_PRL, Loudon_2017}.

Fig.~\ref{Fig-1} shows evolution of skyrmion cluster in a  FeGe nanostripe fabricated from bulk crystal with the method described in~\cite{Jin_17}. {The characterization of such samples with Electron Energy Loss Spectroscopy shows that typical thickness variation does not excised 3\%~\cite{Wang_2017}. }
At low field the skyrmions preferentially assemble near the edges and with increasing field they migrate to the middle part of the stripe without noticeable lost of their long-range order, compare Fig.~\ref{Fig-1}(a) and (b).
One has to note that an assumption of the presence of the strong pining centers which in general can be considered as an alternative explanation for skyrmion clustering, in fact, is not supported by our observations which clearly show a high mobility of the SkTs (Fig.~\ref{Fig-1}(b),(c)) under the action of varied magnetc field even at relatively low temperature $T\sim100$~K. 
Moreover, in Section~\ref{Jumps} we describe another interesting phenomenon of spontaneous leaps of single skyrmion between the two positions, indicating a high mobility of SkTs activated by elevating the temperature to 150~K (see also~\cite{suppl}).
At high $B_\mathrm{ext}$\,=\,484~mT, the long-range order in skyrmion cluster is violated while the skyrmions aggregate near the middle line of the stripe (Fig.~\ref{Fig-1}(c)).

\begin{Figure}
\centering
\includegraphics[width=7cm]{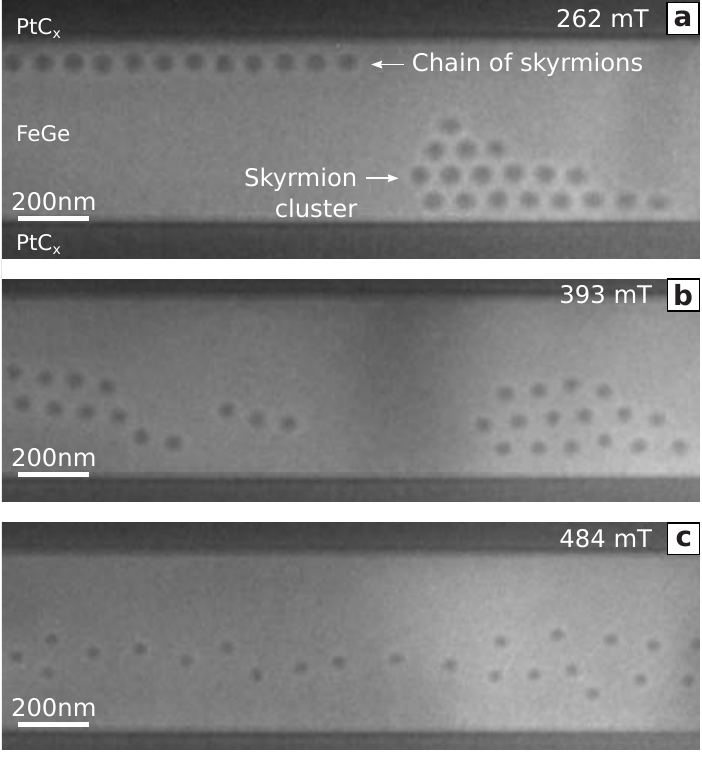}
\caption{\small
Lorentz TEM images showing the evolution of skyrmions arrangement in a wide,~$\sim\,500$~nm, FeGe nanostripe with thickness~$\sim\,120$~nm with increasing magnetic field applied normany to the plate.
(a) Skyrmion chain and skyrmion cluster with long range order formed near the edges of the sample at low field.
(b) Skyrmion cluster migrating to the central part of the sample with increasing field. A long range order in skyrmions arrangement and the distance between skyrmions is conserved.
(c) Disordered skyrmion cluster in the middle part of the nanostripe at high magnetic field. The distance between skyrmions is not conserved.
All Lorentz-TEM images are taken at $T\,=\,100$~K in underfocus conditions with a focused value~$\sim\,300$~$\mu$m. 
}
\label{Fig-1}
\end{Figure}

On a qualitative level, the attractive skyrmion-skyrmion interaction can be a good explanation for these observations at low field and the violation of the long-range order at high field may serve as an indicator that the character of the inter-skyrmion as well as skyrmion-edge interaction changes from attraction to repulsion.
Nevertheless, it is well known that even purely repulsive particles can form the clusters and clumps~\cite{Klein,Glaser}.
Thereby, the clustering itself is not sufficient argument in favor of the hypothesis of long-range inter-skyrmion attraction.

\begin{figure*}
\centering
\includegraphics[width=17cm]{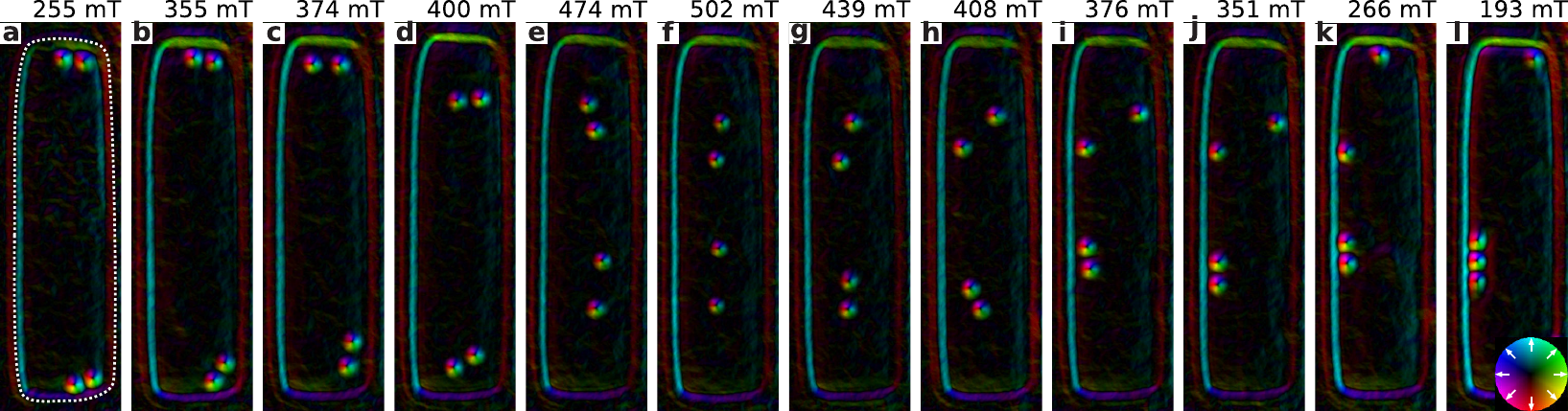}
\caption{\small 
Field-driven evolution of interacting  pairs of skyrmions in a nearly rectangular FeGe plate (the boundaries are marked in panel (a) with white dotted lines) with the width, length, and thickness of 430 nm, 1590 nm, and 120 nm, respectively.
Color represent direction of in-plain component of magnetization, see inset in (l). 
Attractive skyrmion-skyrmion and skyrmion-edge interactions are observed at low magnetic field, while at high magnetic field character of the interactions changes to repulsive. Increasing (a)-(f) and decreasing (g)-(l) field magnetization process illustrate reversibility of the states with very weak hysteresis effect. 
The lower-left isolated skyrmion and corresponding skyrmions pair are chosen to estimate the inter-skyrmion and skyrmion-edge distances, respectively, with the date presented in Fig.~\ref{Fig-3}(c)-(d).
}
\label{Fig-2}
\end{figure*}

In order to reveal the genuine character of skyrmion interactions one has to go beyond the collective phenomena and study the isolated skyrmion pairs. 
Besides that an adequate theoretical model for skyrmion interactions must be in a quantitative agreement with experimental data which in turn requires taking into account the effects of stray field (dipole-dipole interaction) which is always naturally present. 
Such long-range interaction is known to be significant in such a systems~\cite{Shibata_2017}, but was ignored in earlier studies on skyrmion interactions~\cite{Leonov_16, Leonov_APL, Muller_PRL, Loudon_2017}. 
In the following, we present a complementary experimental and theoretical results unambiguously revealing the details of the inter-skyrmion and skyrmion-edge interactions.

Fig.~\ref{Fig-2} shows field-driven evolution of two pairs of SkTs at increasing (a)-(f) and decreasing $B_\mathrm{ext}$ (g)-(l) in the nanostripe.
We measured the field dependence of equilibrium inter-skyrmion and skyrmion-edge  distances and identified the critical value of $B_\mathrm{ext}$, above which the SkTs start to weakly repel each other.
The initial state with reduced number of SkTs, Fig.~\ref{Fig-2}(a) has been achieved via the control of the number of helical spirals in the nanostripe which in turn can be adjusted during the demagnetization process.
As has been revealed in our earlier experiments, at low temperatures, each period of the helix converges to one single SkT by increasing $B_\mathrm{ext}$~\cite{Du_15}.
At $B_\mathrm{ext}\!=\!255$~mT, two pairs of skyrmions are positioned near the edge of the sample on the large distance, thus the interaction between pairs is negligible.
Up to the magnetic field $\sim 400$~mT, the nearest skyrmion-skyrmion distance, $d_\mathrm{ss}$, remains almost unchanged, (a)-(d),
but in the range from 400~mT to 500~mT, it increases abruptly, (e)-(f). 
In contrast to $d_\mathrm{ss}$, the skyrmion-edge distance, $d_\mathrm{se}$, increases gradually at small $B_\mathrm{ext}$ (a)-(c), and start to increases abruptly above 350~mT. 
At a higher $B_\mathrm{ext}\!\sim\!500$~mT, the skyrmions are distributed in the middle of the nanostripe and show very weak variation in their positions with farther increasing field up to the collapse.
The observed behavior clearly demonstrates that the character of inter-skyrmion interaction can be switched from strong attractive at low field to weakly repulsive at high $B_\mathrm{ext}$. 
The above-discussed behavior of skyrmions is well reproduced (a complete field-driven motion of skyrmion is shown in Video\,2~\cite{suppl}), and is almost unchanged when the sample is tilted with respect to the plane normal up to 5$^\circ$ (see Section~\ref{TiltedField}).

An important feature of the field-driven evolution of inert-skyrmion and skyrmion-edge distances is its fully reversibility.
Otherwise, possible explanation would be the presence of the pinning centers which obstruct skyrmions to run away from one another.
With decreasing $B_\mathrm{ext}$ down to 408~mT (Fig.~\ref{Fig-2}(h)), the $d_\mathrm{ss}$ of the bottom pair is significantly decreased which indicates the appearance of attractive interaction again.
It is seen that the top pair of particles has been severed due to the attraction acting on each skyrmion from the side of the opposite edges of the sample. 
With further decreasing of the field, all SkTs move close to the edge and form either paired state or remain isolated (i)-(k). 
It is worth paying attention to the relatively large distance between single SkT and pair of SkTs on the left side of the sample at 266~mT (Fig.~\ref{Fig-2}(k)). 
Despite the extremely large distance between them, these three skyrmions form a chain when the $B_\mathrm{ext}$ further reduces down to 193~mT (Fig.~\ref{Fig-2}(l)), suggesting that SkTs are able to attract each other on the distances  much larger than  the characteristic size of the skyrmions, $L_\mathrm{D}$.

\begin{figure*}
\centering
\includegraphics[width=16cm]{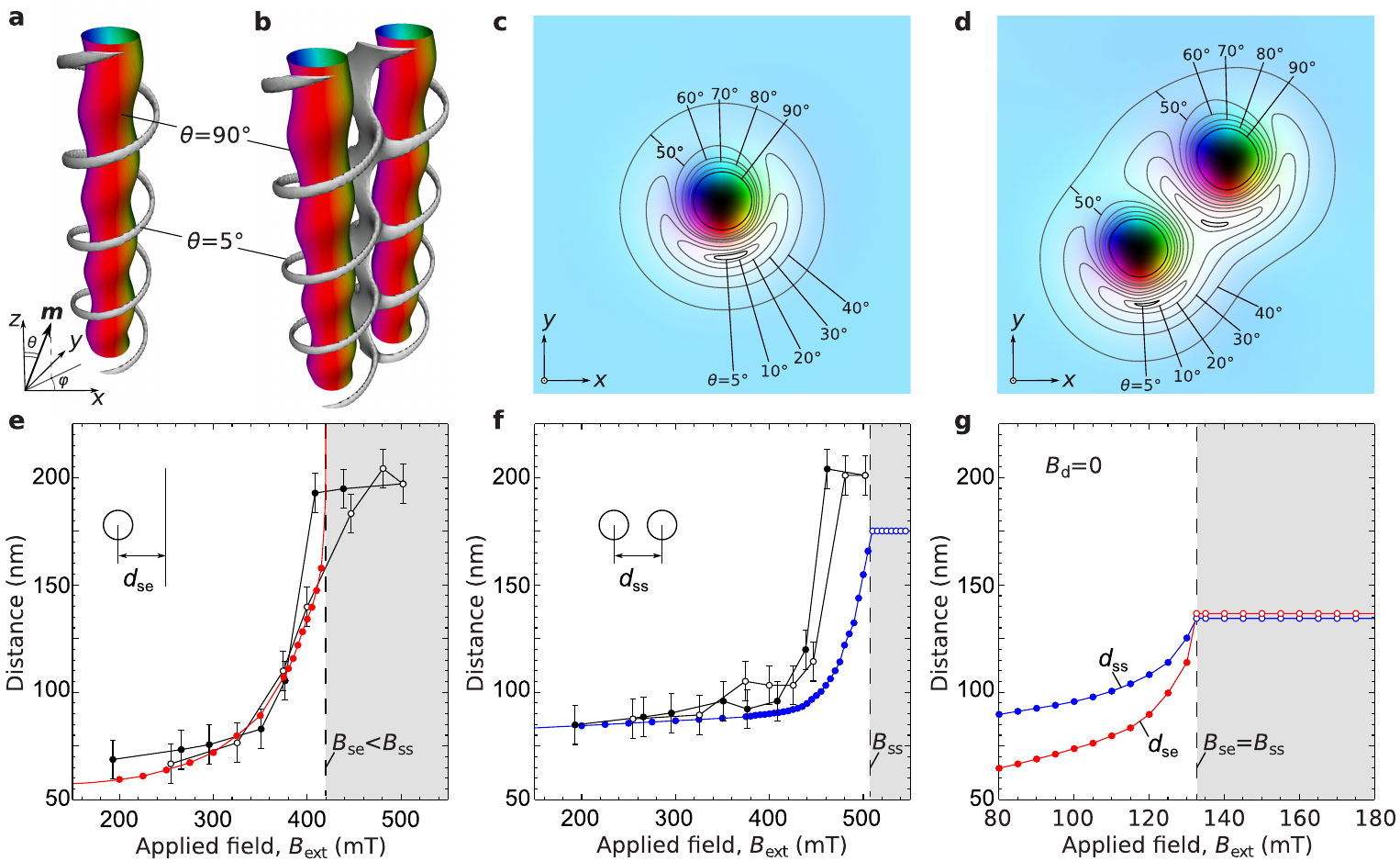}
\caption{\small
(a) and (b) represent the isosorfaces $\theta\!=\!90^\circ$ and $\theta\!=\!5^\circ$ for typical segment of single SkT and coupled pair of SkTs respectively while for finite thickness film the corresponding SkTs characterized by distortion near the ends of the tube due to chiral surface twist~\cite{Rybakov_13}. 
The standard color code indicate the direction of the magnetic moments on a unit sphere where black and white corresponds to $m_z\!=\!-1$ ($\theta\!=\!180^\circ$) and $m_z\!=\!+1$ ($\theta\!=\!0^\circ$) respectively. 
(c) and (d) represent cross section of SkT and pair of SkTs shown in (a) and (b) with $z\!=\!\mathrm{const}$ plain. Closed curves indicate isolines for $\theta$ angle.
(e) Skyrmion-edge distance as function of filed.
(f) Skyrmion-skyrmion distance.
Open and solid (black) circles in (c) and (d) correspond to increasing and decreasing field respectively.
Red and blue circles represent theoretical dependencies.
(g) The theoretical dependencies for $d_\mathrm{se}$ and $d_\mathrm{ss}$ calculated in assumption that demagnetizing field inside the sample, $B_\mathrm{d}\!=\!0$.} 
\label{Fig-3}
\end{figure*}

This attractive interactions can be well understood in the following way.
At low magnetic field, the isolated skyrmion is surrounded by the cone phase, giving rise to nontrivial 3D SkT (Fig.~\ref{Fig-3}(a)).
When two of such inhomogeneous SkTs approach each other, the mutual volume of the pair turns to be smaller than the total volume of two isolated SkTs (Fig.~\ref{Fig-3}(b)).
Accordingly, the total energy of such coupled state with respect to the energy of the cone phase becomes lower and make the coupled SkTs being energetically more favorable than two isolated SkTs. 
However, too small distance between two SkTs leads to an additional distortion of the spins configuration of the two SkTs, which are energetically unfavorable.
The competitions of this two effects: i) reduction of the total volume occupied by the coupled two SkTs and ii) distortion of isolated SkTs, results in an equilibrium inter-skyrmion distance with Lennard-Jones like potential of interparticle interaction~\cite{Leonov_16}. 
Moreover, a strict mathematical analysis of the basic model of isotropic chiral magnet revealed that the spin texture of isolated SkT appearing as an excitation on the background of the conical phase has hidden symmetry. 
The later means that without any loss of generality the exact solution for isolated translationally invariant SkT, as shown in Fig.~\ref{Fig-3}(a), can be obtained via special 2D model Hamiltonian, for details see Section~\ref{BasicModel}. 
This solution corresponds to a 2D ``oblique'' skyrmion (similar to the texture of the pair of half lumps in easy plane baby Skyrme model~\cite{JS_2010} and skyrmions in tilted magnetic field~\cite{Lin,Leonov_Neel_sk}), as shown in Fig.~\ref{Fig-3}(c), in ferromagnetic background inclined by angle $\theta_\mathrm{cone}$ with respect to $\mathbf{B}_\mathrm{ext}$ corresponding to equilibrium angle of conical phase. 
The complete spin texture of 3D SkT is then followed from analytical mapping based on affine transformation. 
Although in films of finite thickness the additional twists of magnetization arise near the surfaces~\cite{Rybakov_13}, the underlying part of the tube preserves aforementioned symmetry.
The presence of another SkT in the system in turn breaks this symmetry, as well illustrated by means of overlapping isosurfaces of two SkTs in Fig.~\ref{Fig-3}(b) and (d).
The line which connects the cores can not be preserved under the affine transformation.
Coupled SkTs can be considered as one object which is also a skyrmionic solution with higher topological charge.

To validate experimentally observed behaviors, we performed micromagnetic simulations by means of MuMax3~\cite{MuMax3} for the realistic specimen geometry (see Section~\ref{Micromag}). 
The theoretical curves in Fig.~\ref{Fig-3}(e),(f) are in a very good qualitative and quantitative agreement with experimental data. 
For the critical fields $B_\mathrm{se}\!\sim\!420$~mT and $B_\mathrm{ss}\!\sim510\!$~mT the corresponding theoretical dependencies for $d_\mathrm{se}$ and $d_\mathrm{ss}$ tends to infinity which reflects the change in the character of interactions from attractive to repulsive. 
It happens because at strong $B_\mathrm{ext}$ the cone phase reaches the saturation which means that the ``vacuum'' surrounding SkTs become field-polarized and isotropic. 
The SkTs in such ferromagnetic background possessing radially symmetric structure  behave identically to skyrmions in 2D system and always repel each other~\cite{Bogdanov_1995}. 
Thereby, the measured value of critical field $B_\mathrm{ss}$ also can be thought as the saturation field for conical phase.

It is important to note the difference between the values of critical fields $B_\mathrm{se}$ and $B_\mathrm{ss}$.
As follows from experimental and theoretical dependencies presented in Fig.~\ref{Fig-3}(e),(f) $B_\mathrm{se}\!<\!B_\mathrm{ss}$ meaning that skyrmion-edge interaction changes its character to repulsive at significantly lower field compare to inter-skyrmion interaction.
For $B_\mathrm{se}\!<\!B_\mathrm{ext}\!<\!B_\mathrm{ss}$ SkTs form clusters far from the edges in the center of the sample. 
This agrees with recently observed effect of skyrmion clustering in the center of the FeGe plate~\cite{Zheng_17}. 
In order to reveal the nature of this effect we performed comparative calculation for $d_\mathrm{se}$ and $d_\mathrm{ss}$ where we ignore the energy contribution of demagnetizing field (Fig.~\ref{Fig-3}(g)). 
Besides the significant reduction of absolute values of  $B_\mathrm{se}$ and $B_\mathrm{ss}$, which reflect important role of demagnetizing fields for quantitative agreement one can clearly see that the dependencies for $d_\mathrm{se}$ and $d_\mathrm{ss}$ tend to infinity at the same critical field, $B_\mathrm{se}=B_\mathrm{ss}$.
The latter means that the effect of skyrmion clustering in the center of the sample reflects strong influence of demagnetizing fields.  
Both $d_\mathrm{se}$ and $d_\mathrm{ss}$ exhibit nearly constant value above critical field.
Such saturation occurs because the energy contribution of weak inhomogeneities, thermal fluctuations etc. becomes comparable or even exceeds the repulsive inter-skyrmion forces which are exponentially small at large distances. 
In the simulations the saturation of of $d_\mathrm{se}$ and $d_\mathrm{ss}$ is mainly caused by finite precision of numerical calculations.
All aforementioned effects are also observed in large size samples (see Section~\ref{LargePlate}) and can be well explained in terms of forces of inter-skyrmion and skyrmion-edge interactions and their dependences on external magnetic field provided in Section~\ref{Forces}.

In summary, we present first experimental evidence for Lennard-Jones type interaction between individual skyrmions revealed by direct observation of reversible field-driven evolution of interparticle equilibrium distances. 
In a good agreement with results of micromagnetic simulations, it is shown that with increasing field the character of long-range skyrmion interactions changes from attractive to repulsive. 
Finally, it is worth to emphasize that the revealed mechanism of attractive interaction for 3D skyrmions is completely different from other systems such as for instance frustrated 2D magnets~\cite{Rozsa} or type-1.5 superconductors~\cite{Babaev}. 

\section*{\normalsize Acknowledgments}

The authors would like to thank E.~Babaev for helpful discussions. This work was supported by the National Key R$\&$D Program of China, Grant No. 2017YFA0303201, the Key Research Program of Frontier Sciences, CAS, Grant No. QYZDB-SSW-SLH009, the Natural Science Foundation of China, Grant No. 51622105, 11474290,  {the Key Research Program of the Chinese Academy of Sciences, KJZD-SW-M01, the Major/Innovative Program of Development Foundation of Hefei Center for Physical Science and Technology, Grand No. 2016FXCX001} and the Youth Innovation Promotion Association CAS No. 2015267. The work of F.\,N.\,R. was supported by the Swedish Research Council Grant No. 642-2013-7837 and by G\"{o}ran Gustafsson Foundation for Research in Natural Sciences and Medicine. The research of A.\,B.\,B. was carried out within the state assignment of FASO of Russia (theme ``Quantum'' No. 01201463332).

\vspace{0.5\baselineskip}

\renewcommand{\section}[2]{}

\end{multicols}

\setcounter{equation}{0}
\setcounter{figure}{0}

\renewcommand{\thesection}{\normalsize\text{A}\arabic{section}}
%\numberwithin{equation}{section}
%\numberwithin{figure}{\text{A}}
\renewcommand\theequation{\text{S}\arabic{equation}} 
\renewcommand\thefigure{\text{S}\arabic{figure}}

\part*{\centering \large Appendices}\label{app}

\section{\normalsize Jumps of skyrmion}\label{Jumps}
\begin{figure}[hb]
\minipage{0.42\textwidth}
\centering
\includegraphics[width=1.0\textwidth]{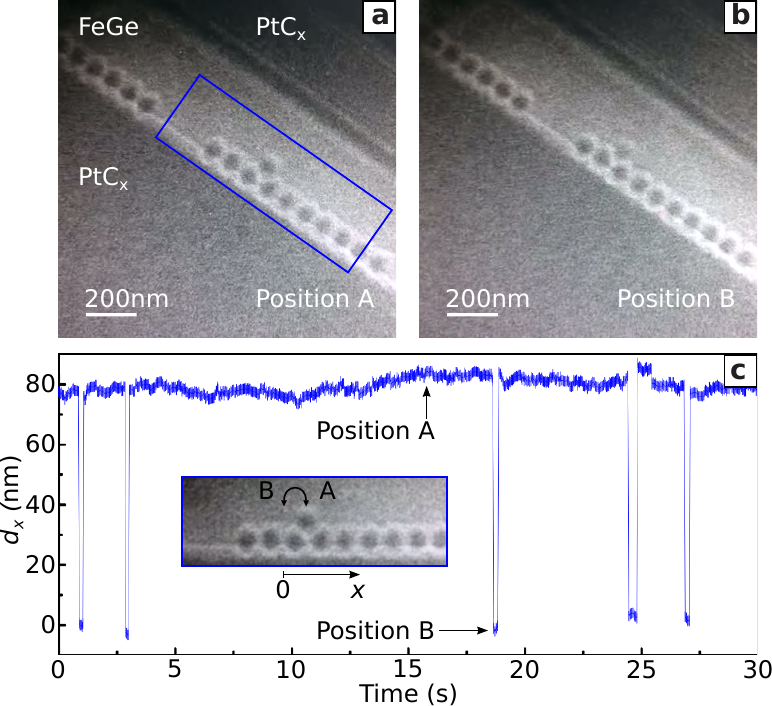}
\caption{\small
(a) and (b) shows skyrmion in two distinct positions along the chain of skyrmions along the stripe edge. 
(c) Time dependence of the position of a jumped skyrmion in a nanostripe with the thickness of~$\sim\,100$~nm at the temperature $T\!\sim\!150$~K and the magnetic field $B_\mathrm{ext}\!=\!235$~mT. 
The inset shows the positions ``A'' and ``B'' between which  skyrmion jumps (see corresponding Video~1~\cite{suppl}). 
} 
\label{Fig-Jumps}
\endminipage\hfill
\minipage{0.54\textwidth}
\centering
\includegraphics[width=1.0\textwidth]{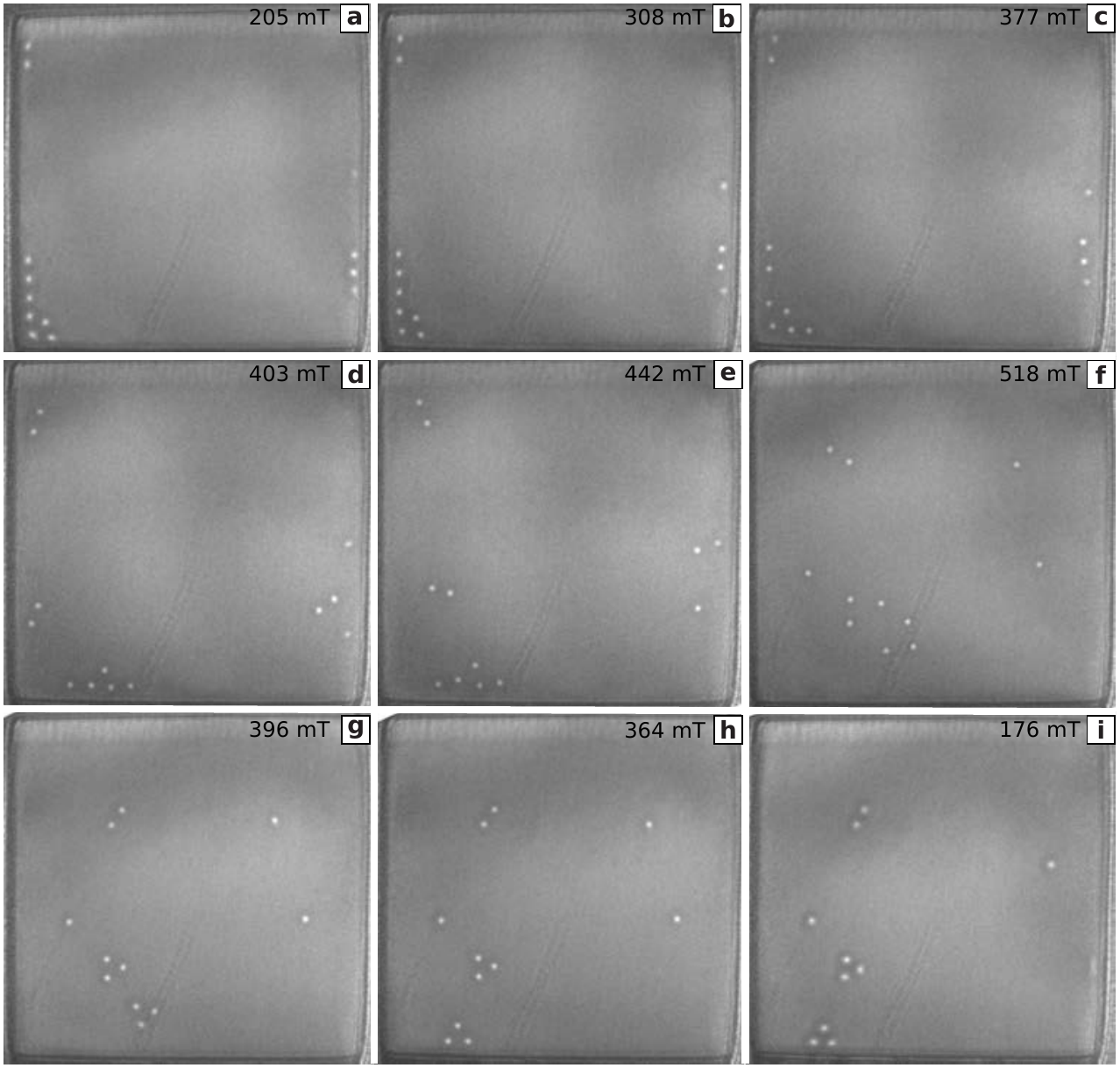}
\caption{\small
Field-driven evolution of skyrmions in a square shape FeGe plate with the thickness~$\sim120$~nm and transversal dimensions of~$\sim\,1.5$~$\mu$m.
(a)-(f) represent images for increasing filed and (g)-(i) are images for decreasing field.
All images have been taken at temperature $T\!=\!100$~K in over-focus regime with defocus distance a focused value~$\sim\,400$~$\mu$m.
} 
\label{Fig-Big}
\endminipage
\end{figure}

Fig.~\ref{Fig-Jumps} illustrates an interesting phenomenon of spontaneous leap of single skyrmion between the two intermediate sites relative to the chain of skyrmions at rest. 
This phenomenon clearly indicates high mobility of SkTs as well as that the skyrmion chain attached to the edge provides potential energy pits for individual SkT moving along such a chain.

\section{\normalsize Large size square plate}\label{LargePlate}
The experiment on large size square plate, Fig.~\ref{Fig-Big}, has been performed with the aim to illustrate few different aspects of the skyrmion interactions problem.

First, it shows reproducibility of the effect of reversal switching of the character of skyrmion interactions with magnetic field for large size samples. It is clearly shows that the equilibrium distance between skyrmions increase gradually with the applied filed, Fig.~\ref{Fig-Big}(a)-(f) and reduces back to initial distances with the reduction of applied field, Fig.~\ref{Fig-Big}(g)-(i).

Second, presented images clearly illustrate effect of saturation of the inter-skyrmion and skyrmion-edge distances and presence of critical distances on which skyrmions may still interact with each other and with the modulations at the edges of the sample. In particular, in ideal system above critical field at which skyrmion-edge interaction acquires a pure repulsive character it is expected thate skyrmions will appear in the center of the domain due to the repulsion actin from each edge of the sample. In real system, however, the center of the domain remain unoccupied by skyrmions, which can be explained by the existence of the critical distance above which the repulsive forces acting on the skyrmions from the edges become negligibly small in comparison to weak thermal fluctuations and different pinning effects.
Another clear evidence of this effect is the fact that some skyrmions which appear at large distances from the edges during the increasing field remains on their positions and can not be attracted by edges because of large distances between them, compare Fig.~\ref{Fig-Big}(a) and (i).

\section{\normalsize Tilted magnetic field}\label{TiltedField}
In Lorentz TEM experiment small angle tilt of the sample is used for  reduction of the diffraction effect. 
Typical value of such tilt angle with respect to plane normal does not exceed 2$^\circ$.
In this regime and up to the tilt angle of~$\sim\,5^\circ$ we have found no significant changes in the skyrmions behavior discussed in the main text. 
Significant difference has been observed only when the tilting angle approaches value of~$\sim\,15^\circ$.
However, this case goes beyond the scope of current work and will be discussed elsewhere.

\section{\normalsize Analysis of the basic model}\label{BasicModel}
Let us consider the basic model of isotropic chiral magnet~\cite{Bar,Bak} defined with the following model Hamiltonian in 3D space:
%W16*2=32W
\begin{align}
\mathcal{H}_0  = \int \Big( \mathcal{E}\{\mathbf{m}\} \Big) \mathrm{d}x\mathrm{d}y\mathrm{d}z, \quad
\mathcal{E}\{\mathbf{m}\} = \mathcal{A} \sum_{i=x,y,z} ( \mathbf{grad}\,m_i )^2 
+ \mathcal{D}\, \mathbf{m}\!\cdot\! [\nabla\! \times \! \mathbf{m}] - M_\mathrm{s} B_\mathrm{ext} m_{z},
\label{E_3D}
\end{align}
%W58
where  $\mathbf{m} \equiv \mathbf{m}(x,y,z)$ is a unit ($\mathbf{m}=\mathbf{M}/M_\mathrm{s}$) continuous vector field, $M_\mathrm{s}$ is the magnetization of the material, $\mathcal{A}$ and $\mathcal{D}$ are the micromagnetic constants for exchange and DMI, respectively. We established that in case of bulk approach both axisymmetric and non-axisymmetric isolated skyrmion can be found as exact solution of the following special 2D Hamiltonian:
%W16
\begin{align}
\mathcal{H}\!=\!\int\!\Bigg(\mathcal{E}\{\mathbf{n}\} - \rho\,(n_x^2\!+\!n_y^2) +  \rho\,\Big(y\frac{\partial\mathbf{n}}{\partial x} - x\frac{\partial\mathbf{n}}{\partial y}\Big)^2\Bigg) \mathrm{d}x\mathrm{d}y,
\label{E_2D}
\end{align}
where $\rho\!=\!\mathcal{D}^2/(4\mathcal{A})$ and unit vector field $\mathbf{n}\equiv\mathbf{n}(x,y)\!=\!\mathbf{m}(x,y,0)$. The texture $\mathbf{n}$ reproduce field $\mathbf{m}$ by coordinated global and local rotations dependent on $z$-coordinate:
%W16*4=64W
\begin{align}
&\mathbf{m}(x,y,z) = 
\begin{pmatrix}
 \cos(k z) & -\sin(k z) & 0 \\
 \sin(k z) & \cos(k z) & 0 \\
 0 & 0 & 1
\end{pmatrix} \mathbf{n}( x^\prime, y^\prime ), \label{symmetry} \\
&x^\prime\!=\!x \cos(k z) + y \sin(k z),\quad y^\prime\!=\!y \cos(k z) - x \sin(k z), \nonumber
\end{align}
where $k=2\pi/L_\mathrm{D}$, cone period $L_\mathrm{D}=4\pi \mathcal{A}/\mathcal{D}$~\cite{Bogdanov_11}. For the values of $B_\mathrm{ext}$ less than saturation field $B_\mathrm{D}\!=\!\mathcal{D}^2/(2M_s\mathcal{A})$~\cite{Bogdanov_11} variational equations for (\ref{E_2D}) gives trivial solution $n_z\!=\!B_\mathrm{ext}/B_\mathrm{D}\!=\!\cos(\theta_\mathrm{cone})$ describing the cone phase.

The spin texture of isolated SkT appearing as an excitation on background of such conical phase has hidden symmetry (\ref{symmetry}). The presence of another SkT in the system breaks this symmetry because the line, which connects the cores of skyrmions can not be preserved under the affine transformation underlying Eq. (\ref{symmetry}).

\section{\normalsize Micromagnetic simulations}\label{Micromag}
\begin{figure}
\centering
\includegraphics[width=16cm]{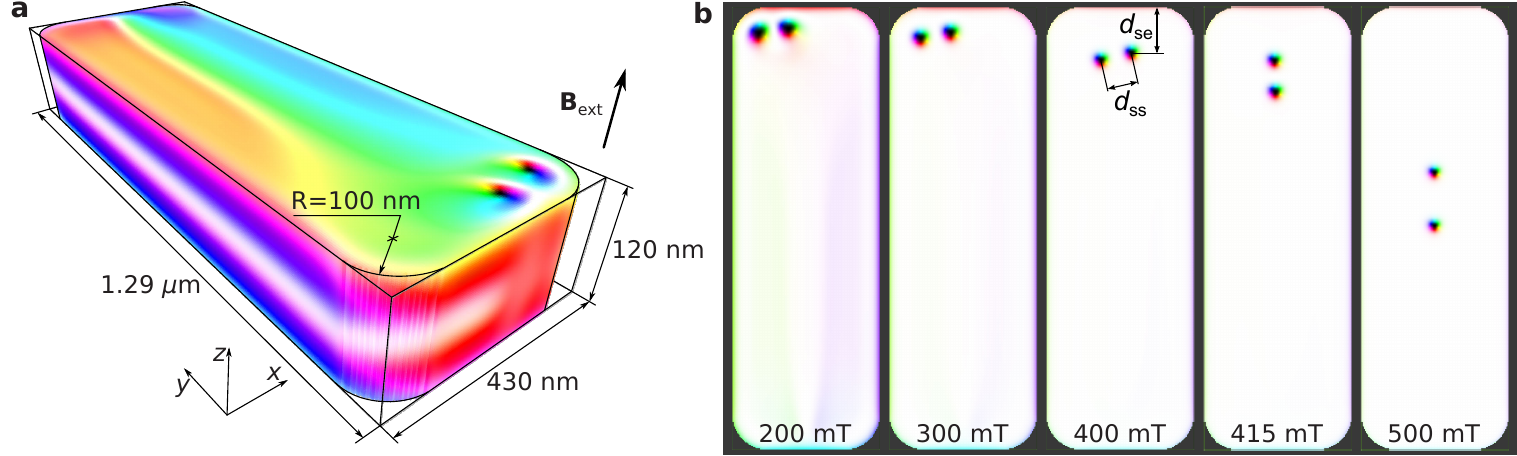}
\caption{\small 
(a) Size and shape of simulated domain with color code showing magnetization distribution in initial state at $B_\mathrm{ext}\!=\!200$~mT.
(b) Equilibrium magnetization distribution averaged over the thickness of the plate shown for different applied fields.
} 
\label{Fig-Sup}
\end{figure}

In nano structured samples the interaction of magnetization with demagnetizing fields $\mathbf{B}_\mathrm{d}$ generated by divergence of magnetization at the boundaries and in the volume of the sample became important for quantitative description~\cite{Shibata_2017}.
In the simulations, we consider the following energy functional:
\begin{eqnarray}
E\!=\!\int_V \Big(  
\mathcal{A} \sum_{i=x,y,z} ( \mathbf{grad}\,m_i )^2 
+ \mathcal{D}\, \mathbf{m}\!\cdot\! [\nabla\! \times \! \mathbf{m}] \Big.
\Big. - M_\mathrm{s} B_\mathrm{ext} m_{z} - 
\frac{1}{2}\, M_\mathrm{s}\, \mathbf{B}_\mathrm{d} \cdot \mathbf{m}  
\Big)\mathrm{d}x\mathrm{d}y\mathrm{d}z.
\label{E_tot}
\end{eqnarray} 
We use typical value for saturation magnetization for FeGe $M_\mathrm{s}\!=\!384$~kAm$^{-1}$ and the DMI constant defined via equation for equilibrium period of helicoid~\cite{helix,Bogdanov_11,Beg_15}: $\mathcal{D}\!=\!4\pi\mathcal{A}/L_\mathrm{D}$. 
The exchange stiffness constant $\mathcal{A}$ has been found from the fit of experimentally measured dependencies for $d_\mathrm{se}$ and $d_\mathrm{ss}$. 
We changed $\mathcal{A}$ in range $[2.0 - 9.0]$~pJm$^{-1}$ with the step of 0.25~pJm$^{-1}$.
We have found the best fit to experimental dependencies with $\mathcal{A}\!=\!3.25$~pJm$^{-1}$.
The mesh of the simulated domain was 128\,$\times$\,384\,$\times$\,64 cells.
Fig.~\ref{Fig-Sup}(a) show an initial state at $B_\mathrm{ext}\!=\!200$~mT, which was archived by energy minimization of homogeneously magnetized plate with imposed pair of SkTs. At each value of $B_\mathrm{ext}$ varying from 200~mT to 600~mT with the step of 25~mT, full energy minimization with a conjugate gradient method has been performed.

\section{\normalsize Estimation of conservative and restoring forces}\label{Forces}

\begin{figure*}[ht]
\centering
\includegraphics[width=9cm]{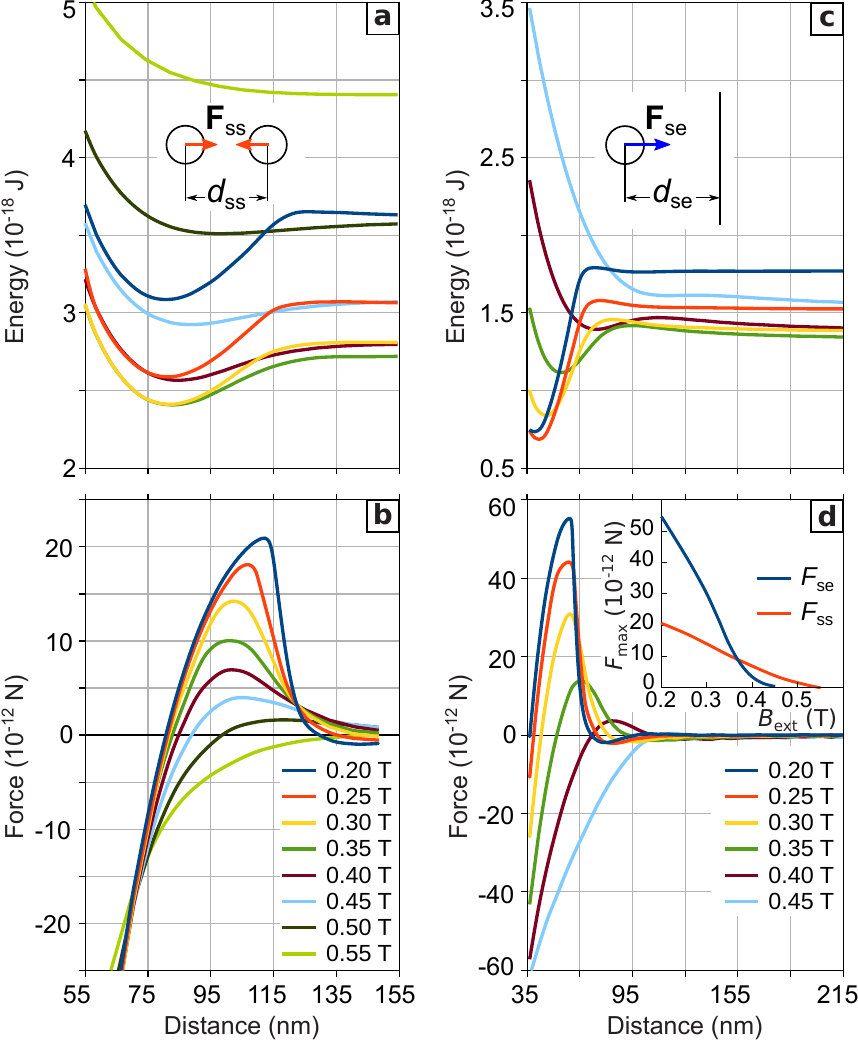}
\caption{\small
(a) Potential of inter-skyrmion interaction energy as function the distance between skyrmions, $d_\mathrm{ss}$ calculated for different values of applied magnetic field. (b) Inter-skyrmion force, $F_\mathrm{ss}=\partial E/ \partial d_\mathrm{ss}$ calculated from the potential, $E(d_\mathrm{ss})$ shown in (a). $F_\mathrm{ss}>0$ ($F_\mathrm{ss}<0$) corresponds to attraction (repulsion) between skyrmions.  
(c) Potential of skyrmion-edge interaction energy as function the distance between skyrmion and the edge of the stripe, $d_\mathrm{se}$ for different applied field. (d) Skyrmion-edge force, $F_\mathrm{se}=\partial E/ \partial d_\mathrm{se}$ calculated from the potential, $E(d_\mathrm{se})$ shown in (c).
Inset in (d) shows dependence of the maximal value of attraction forces $F_\mathrm{ss}$ and $F_\mathrm{se}$ as function of applied field.
} 
\label{Fig-potencial}
\end{figure*}

From simulations performed with MuMax3 we estimate that binding energy $U$ of skyrmions (i.e. the difference between the total energy of two isolated skyrmions and energy of their bound state) in the first approximation has linear depends on the applied field:
\begin{align}
\frac{\partial U } {\partial B_\mathrm{ext}} \approx -k,
\end{align}
where $k \approx 7\times 10^{-18}$~J\,T$^{-1}$. The same is true for skyrmion-edge binding energy and calculations gives approximately the same value for the coefficient $k$. The corresponding conservative force is 
\begin{align}
F = - \frac{\partial U } {\partial d} = 
- \frac{\partial U } {\partial B_\mathrm{ext}} \left( \frac{\partial d} {{\partial B_\mathrm{ext}}} \right)^{-1} 
\approx 
k \left( \frac{\partial d} {{\partial B_\mathrm{ext}}} \right)^{-1} .
\end{align}
One simple mechanical analogy explaining the physical meaning of conservative force is vertical spring pendulum, where the analog of force $F$ is the tension of the spring in an equilibrium state and the analog of external magnetic field $B_\mathrm{ext}$ is gravitation.

Comparing the slopes for dependencies in Fig.~\ref{Fig-3}(e),(f) for fixed $B_\mathrm{ext}$ values one can see that at high magnetic field inter-skyrmion force is stronger than that for skyrmion-edge interaction while the order of magnitude for both forces is~$\sim\, 10^{-11}$~N.

Alternative approach to estimate skyrmion interaction forces from above is based on calculation of maximum forces. For that, first we found approximate potential for the inter-skyrmion and skyrmion-edge energies following the approach used in~\cite{Leonov_16}, but applied it to micromagnetic functional (\ref{E_tot}) taking into account stray field energy term and materials parameters given in Section~\ref{Micromag}.
All calculations discussed below has been performed for very wide plate of chiral magnet with the size of $1\mu$m$\times 1\mu$m, thickness 120~nm and discretization mesh of 256\,$\times$\,256\,$\times$\,64 cells.
The periodical boundary conditions have been set along $x$-axis only for skyrmion-edge interaction and along both $x$- and $y$-axis for inter-skyrmion interaction. An external magnetic field is pointing along positive direction of $z$-axis.
In order to fix the position of skyrmions the magnetization of certain cuboids of discretized mesh coinciding with the main axis of skyrmion tube has been excluded from minimization via the MuMax3 build-in function ``frozenspins()''.
For each fixed magnetic field we were changing the position of skyrmions in initial state and position of fixed magnetization cuboids respectively and then the full energy minimization were performed for each of configurations.
The energies of the skyrmion pair as function of the distance between skyrmions is shown in Fig.~\ref{Fig-potencial}(a) for different values of applied field.
The energy of a single skyrmion as function of the distance to the free edge of the stripe is shown in Fig.~\ref{Fig-potencial}(c).
The corresponding forces restoring skyrmion to its equilibrium position ($F\,=\,0$) are shown in Fig.~\ref{Fig-potencial}(b) and (d).
Maximal value of such inter-skyrmion and skyrmion-edge forces are shown in inset of Fig.~\ref{Fig-potencial}(d).
Such estimation of the forces acting on the skyrmion are in a good agreement with estimation of conservative forces discuses above in this section.
As follows from the dependencies $F_\mathrm{max}(B_\mathrm{ext})$, the skyrmion-edge force $F_\mathrm{se}$ dominates over inter-skyrmion forces $F_\mathrm{ss}$ at low magnetic field.
In a good agreement with the conclusions given in the main text of the manuscript $F_\mathrm{se}$ converges to zero and change its sign at field~$\sim\,0.45$~T while $F_\mathrm{ss}$ change its sign (possessing repulsive character) at slightly higher field~$\sim\,0.55$~T.
One has to emphasize that the direct comparison of absolute values of conservative forces discussed above and maximal restoring forces calculated from the energy profiles, strictly speaking, has no meaning because two type of forces have different physical meaning.
Moreover, due to artificial distortions introduced by fixed (frozen) magnetization in certain positions of the simulated domain the  estimate of restoring forces has much less precision.
On the other hand, a precise estimate of conservative forces requires double precision (64 bits) of numerical solver, while publicly available version of MuMax3 at the moment allows the precision only up to single-precision floating-point format (32 bits).

\renewcommand{\section}[2]{}

\end{document}